\documentclass[aps,prb,twocolumn,superscriptaddress, showpacs]{revtex4-1}
\usepackage{graphicx}
\usepackage{dcolumn}
\usepackage{bm}
\usepackage{lipsum}
\usepackage{amsmath}
\begin{document}
\title {Defect states in LiFeAs as seen by low temperature scanning tunneling microscopy and spectroscopy}
\author{R. Schlegel}
\author{P. K. Nag}
\author{D. Baumann}
\author{R. Beck}
\affiliation{Leibniz-Institute for Solid State and Materials Research, IFW-Dresden, 01069 Dresden, Germany}

\author{S. Wurmehl}
\affiliation{Leibniz-Institute for Solid State and Materials Research, IFW-Dresden, 01069 Dresden, Germany}
\affiliation{Institute for Solid State Physics, TU Dresden, 01069 Dresden} 

\author{B. B\"uchner}
\affiliation{Leibniz-Institute for Solid State and Materials Research, IFW-Dresden, 01069 Dresden, Germany}
\affiliation{Institute for Solid State Physics, TU Dresden, 01069 Dresden} 
\affiliation{Center for Transport and Devices, TU Dresden, 01069 Dresden, Germany}

\author{C. Hess}
\email{c.hess@ifw-dresden.de}
\affiliation{Leibniz-Institute for Solid State and Materials Research, IFW-Dresden, 01069 Dresden, Germany}
\affiliation{Center for Transport and Devices, TU Dresden, 01069 Dresden, Germany}

\date{\today}

\begin{abstract}
 We present a microscopic investigation of frequently observed impurity-induced states in stoichiometric LiFeAs using low temperature scanning tunneling microscopy and spectroscopy (STM/STS). Our data reveal seven distinct well defined defects which are discernible in topographic measurements. Depending on their local topographic symmetry, we are able to assign five defect types to specific lattice sites at the Li, Fe and As positions. The most prominent result is that two different defect types have a remarkably different impact on the superconducting state. 
A specific and quite abundant Fe-defect with $D_2$-symmetry generates significant impurity-induced additional states primarily at positive bias voltage with pronounced peaks in the on-site local density of states (LDOS) at about 4~mV and 12~mV. On the other hand, a $D_4$-symmetric As-defect causes a significantly enhanced LDOS at both positive and negative bias voltages. We expect that these findings  provide fresh input for further experimental and theoretical studies on elucidating the nature of superconductivity in LiFeAs.
\end{abstract}

\maketitle

\section{Introduction}

Among iron-based superconductors (IBS), the compound LiFeAs takes up a special role. Unlike many other IBS, where superconductivity emerges upon doping from an antiferromagnetic spin density wave (SDW) parent state with Fermi surface nesting \cite{Kamihara2008,Luetkens2009,Rotter2008,Sefat2008}, LiFeAs is a stoichiometric superconductor, i.e., superconductivity is present without any doping \cite{Tapp2008}. 
Furthermore, the compound's fermiology is far away from Fermi surface nesting \cite{Borisenko2010,Kordyuk2011,Umezawa2012,Borisenko2012,Knolle2012,Hess2013,Zeng2013}, and accordingly the system seems to be far away from an antiferromagnetic instability. This is further supported by chemical doping experiments which in all cases lead to a suppression of the critical temperature $T_c$ but never to the evolution into an SDW state \cite{Pitcher2010,Aswartham2011a,Wright2013}. 

The superconducting state of LiFeAs is much under debate. Depending on details of the band structure, theoretical studies either suggest prevailing ferromagnetic fluctuations with an instability towards triplet superconductivity \cite{Brydon2011}, an $s_{+-}$-wave superconducting ground state driven by antiferromagnetic fluctuations \cite{Platt2011,Wang2013,Ahn2014a}, or $s_{++}$-wave superconductivity driven by orbital fluctuations \cite{Saito2014}. Experimental results are likewise puzzling: Heat transport \cite{Tanatar2011}, penetration depth \cite{Hashimoto2012}, and Scanning Tunneling Microscopy and Spectroscopy (STM/STS) studies \cite{Chi2012,Grothe2012,Chi2014} report consistency with $s_{+-}$-wave superconductivity, whereas inconsistency with the latter has been concluded in ARPES experiments \cite{Borisenko2012}. 
In addition, evidence for a time-reversal symmetry breaking state, partially depending on details of electronic doping is found in NMR/NQR \cite{Baek2012}, STM/STS \cite{Haenke2012}, and high-field magnetometry measurements \cite{Li2013}. Finally, compelling evidence for multiple superconducting transitions in the superconducting state have been reported from combined AC-susceptibility/NMR measurements \cite{Baek2013} as well as from STM/STS \cite{Nag2016}.

\begin{figure*}[htb]
\includegraphics*[width=0.67\textwidth]{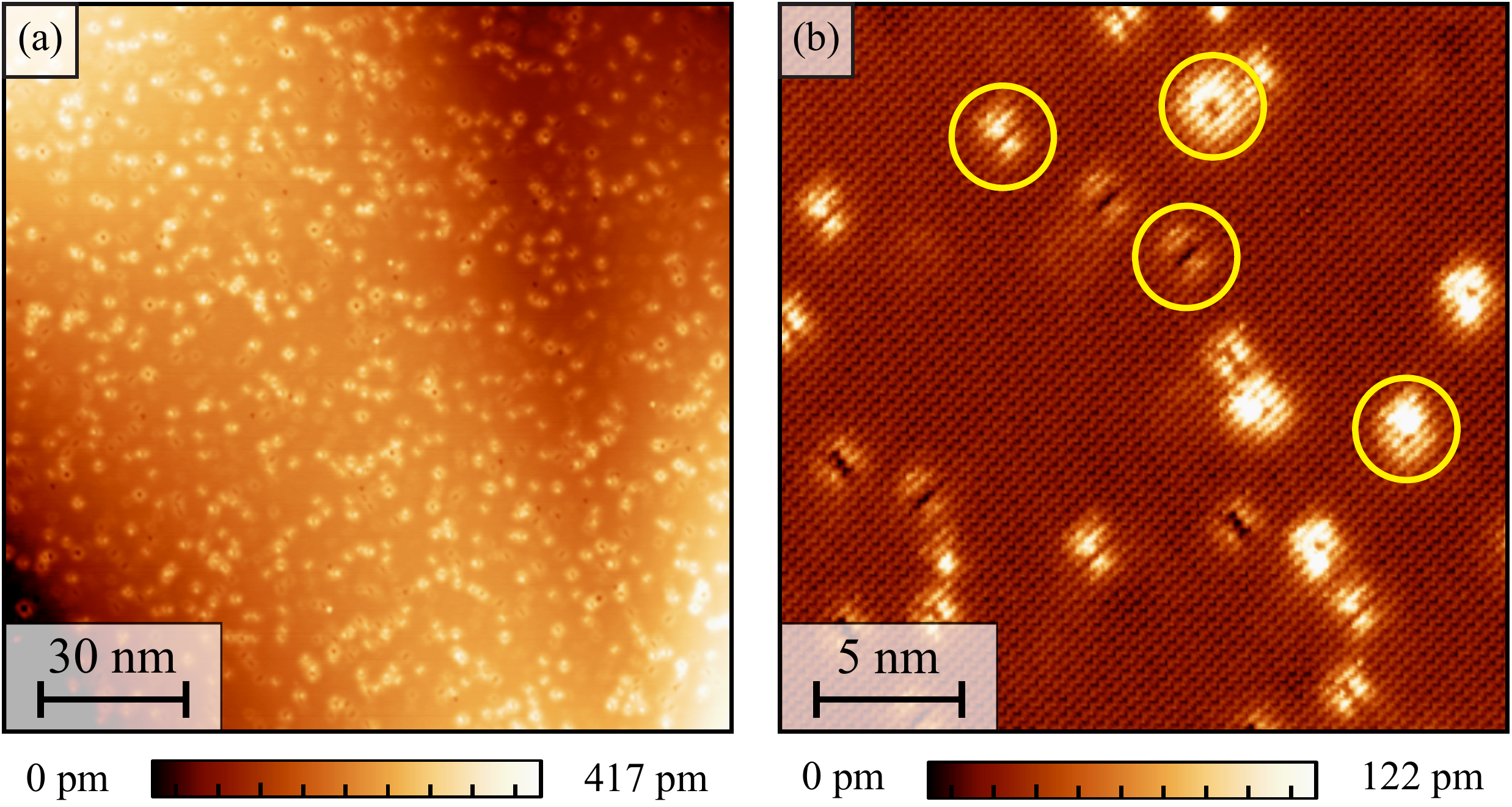}
\caption{(a) Overview topography of 150~nm~$\times$~150~nm ($U_t=-50$~mV, $I_t=100$~pA, $T=5$~K). The surface is very flat without any step edges. Numerous defects appear as bright spots. (b) Zoomed area of 25~nm~$\times$~25~nm ($U_t=-35$~mV, $I_t=100$~pA, $T=5$~K). Four different defect types can be recognized (marked by circles).}
\label{topo_overview}
\end{figure*}

\begin{figure*}[htb]
\includegraphics*[width=0.67\textwidth]{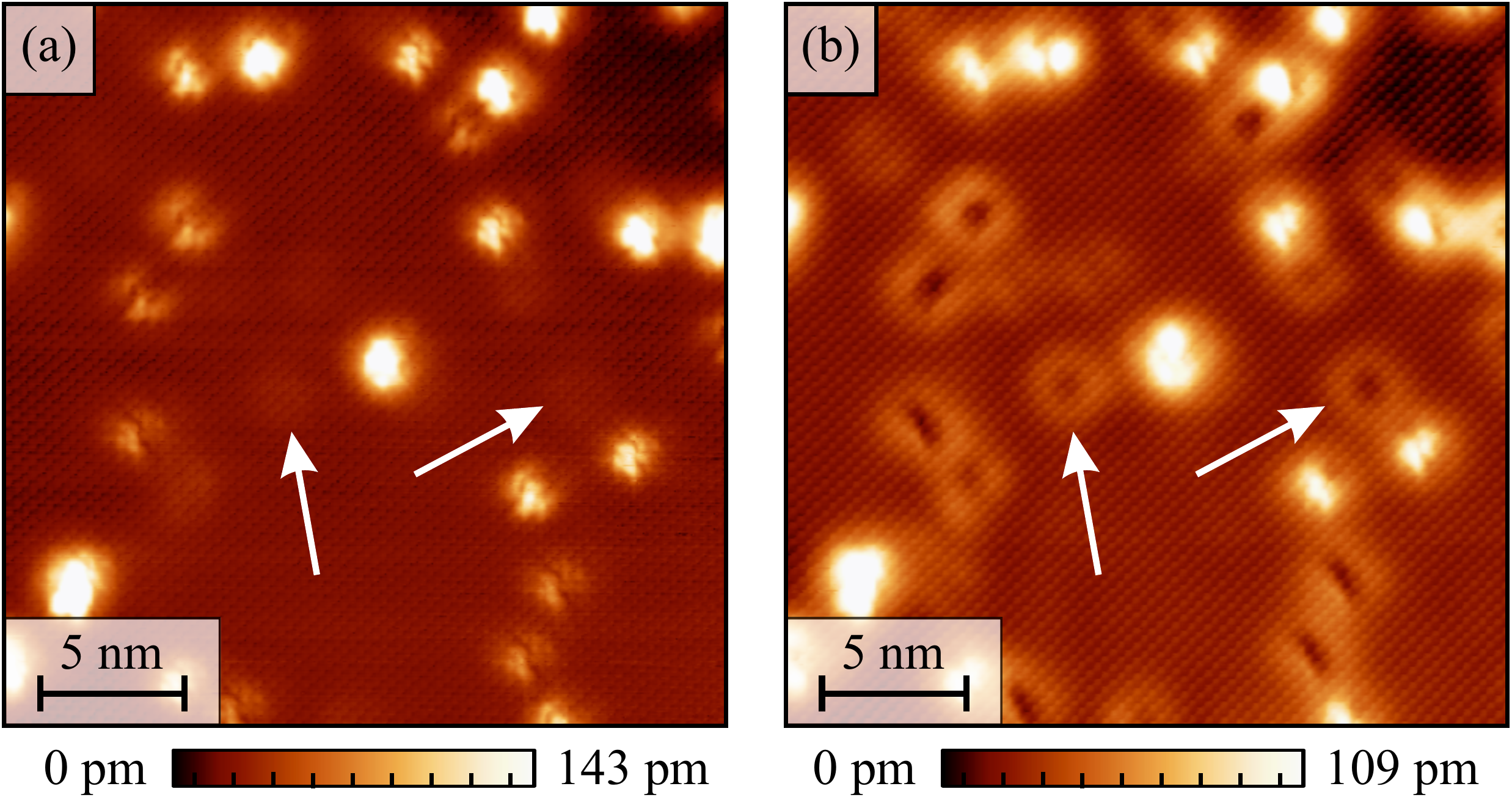}
\caption{Comparison of topographic measurements at positive and negative bias voltage $U_t$ for the same FOV of 25~nm~$\times$~25~nm at $T=5$~K. (a) $U_t=-35$~mV, $I_t=300$~pA. (b) $U_t=35$~mV, $I_t=300$~pA. The appearance of the defects is clearly different in both cases. Furthermore, certain defects (white arrows) are only clearly discernible at positive $U_t$.}
\label{topo_voltage_dependence}
\end{figure*}

The impact of impurities on the superconducting ground state is considered an important approach for probing the superconducting order parameter that roots back to Anderson's theorem on $s$-wave superconductors \cite{Anderson1959}. In particular, depending on the nature of an impurity (also called defect) and the symmetry of the superconducting order parameter these are expected to act as a pair breaker and to give rise to impurity-induced bound states, or not \cite{Balatsky2006}. Unless intentionally doped into a material, impurities generally occur as a natural process during the crystal growth. Thus, probing artificial or natural impurities by STM/STS can potentially provide crucial information about the superconducting state itself \cite{Yazdani1997,Yazdani1999,Pan2000,Hudson2001}. Concerning impurity-induced bound states in IBS theoretical work has addressed both magnetic and non-magnetic impurities \cite{Zhang2009,Tsai2009,Kariyado2010} with the goal to distinguish between $s_{++}$ and $s_{+-}$ order parameters. The results are complicated; while earlier work in simplified two-band models provide information about expected in-gap impurity-induced bound state pairs \cite{Zhang2009,Tsai2009}, a more recent approach acknowledges the multiorbital nature of the band structure and yields distinct suggestions for a bound state, the position of which strongly depends on the assumed impurity potential, and multi-peak structures \cite{Kariyado2010}. Experimentally, Grothe et al. investigated the impurity-induced bound states of LiFeAs by STM/STS experiments on natural impurities \cite{Grothe2012}. Here, we address the same approach towards exploring the superconductivity in this compound. Our results are partially consistent with the earlier study. However, upon exploring a wider energy range that is impacted by the known observed defects at Fe sites and by addressing a so far unreported but quite abundant defect type located at As sites, we observe these to have a radically different, yet strong, impact on superconductivity.

\begin{figure*}[htb]
\includegraphics[width=\textwidth]{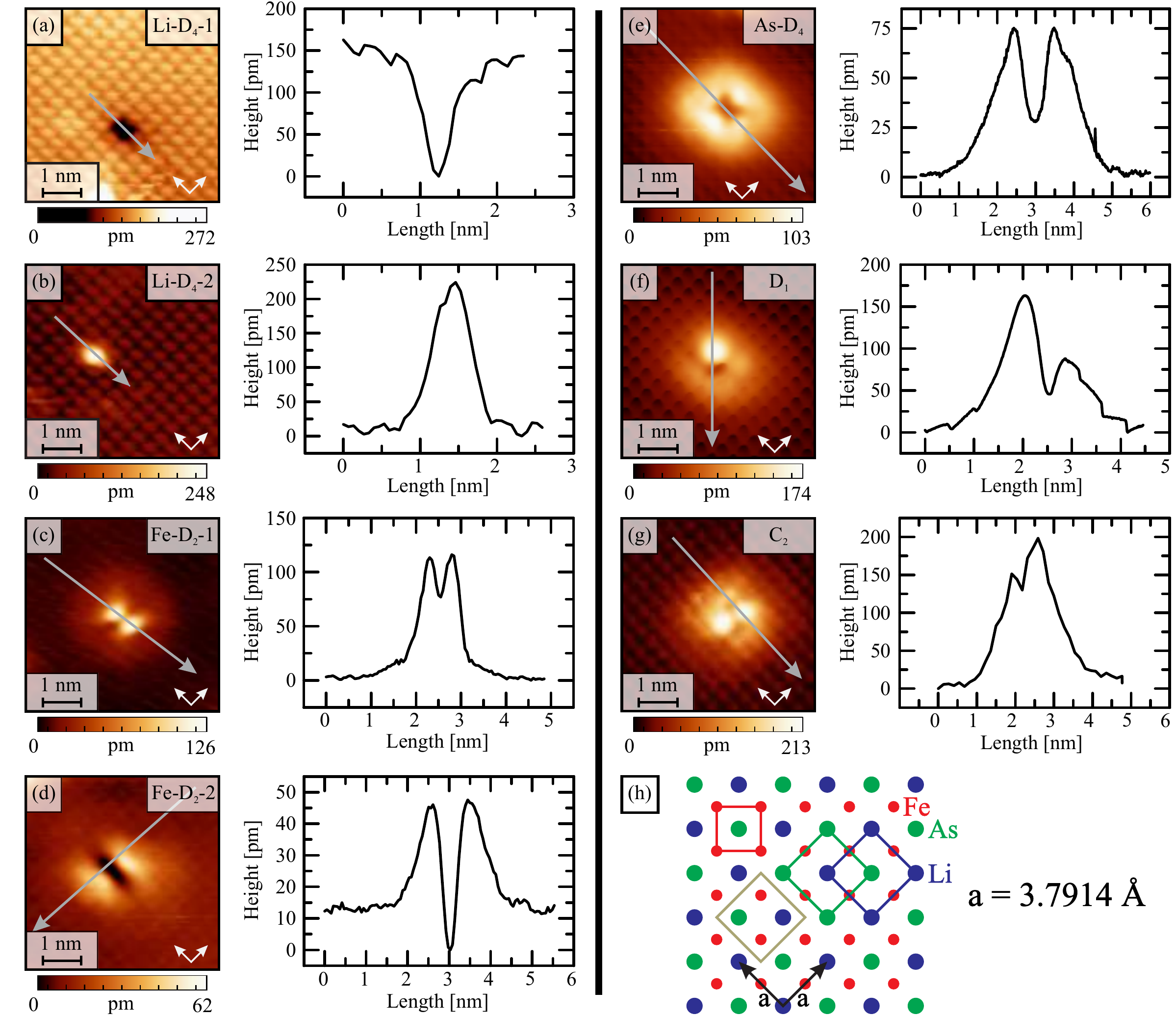}
\caption{Topography images and line profiles of the different observed defects in LiFeAs at 5~K. (a, b) Li vacancy and excess Li,  Li-$D_4$-1 and Li-$D_4$-2 ($U_t=-35$~mV, $I_t=500$~pA), (c, d) defects at Fe-sites Fe-$D_2$-1 and Fe-$D_2$-2 ($U_t=-35$~mV, $I_t=400$~pA), (e) As-defects As-$D_4$ ($U_t=-35$~mV, $I_t=800$~pA), (f, g) low-symmetry defects $D_1$ ($U_t=-35$~mV, $I_t=300$~pA) and $C_2$ ($U_t=-35$~mV, $I_t=300$~pA). (h) Sketch of the $c$-axis projection of the LiFeAs structure.}
\label{all_types_of_defects}
\end{figure*}

\section{Experimental Details}
High quality single crystals of stoichiometric LiFeAs ($\textit{T}_c$ $\approx$ 16 K) have been grown by self flux method as described in \cite{Morozov2010}. A special glove box with long extension has been used for mounting samples into our low-temperature STM inside Ar atmosphere. All STM/STS measurements have been performed in a home-built variable temperature ``dip-stick'' STM \cite{Schlegel2014} which is working between 5~K and room temperature. Electrochemically etched tungsten (W) tips were used for all measurements. 
The samples were cleaved at base temperature (about 5~K) at cryogenic vacuum to obtain fresh and clean surfaces for STM measurements. An external lock-in amplifier was used to record all $dI/dU$ maps as well as spectra with a modulation of 0.4~mV~rms and 1.1111~kHz frequency. All images have been processed using the WSxM software \cite{Horcas2007}.

\section{Results}

\subsection{Topographic overview}
Figure~\ref{topo_overview}a shows results of a topographic overview measurement which reveals an atomically flat cleaved surface without any step edge. Randomly distributed defects can be recognized as spots with bright contrast. A higher resolved topographic image with a field of view (FOV) of 25~nm~$\times$~25~nm shown in Fig.~\ref{topo_overview}b allows to discern the most abundant four different defect types (see below for a detailed classification of the defects). A first comparison with already published data \cite{Chi2012,Grothe2012,Haenke2012,Hanaguri2012,Allan2012} yields a good agreement with respect to the topographic shape and size of most the defects.

Upon further investigating the energy dependence of the defect appearance we observed that it is voltage-asymmetric, i.e., the size of the defects becomes significantly larger at positive bias voltage $U_t$. This can be inferred from Fig.~\ref{topo_voltage_dependence} which compares a 25~nm~$\times$~25~nm FOV at $U_t=\pm35$~mV. This indicates that the local density of states (LDOS) at positive energies (unoccupied states) is influenced stronger by the defects as compared to negative energies. Furthermore, the data exhibit at positive $U_t$ additional defects. This is obvious from the defect positions in Fig.~\ref{topo_voltage_dependence} which are marked by white arrows. At positive  $U_t$ these are clearly resolved whereas at negative  $U_t$ they are almost invisible. Close inspection of such ``hidden'' defects in Fig.~\ref{topo_voltage_dependence}(b) reveals atomic corrugation  on top of these defects. This suggests these defects to be located not in the topmost layer but further below. These findings imply that for evolved spectroscopic studies of the pristine superconducting state of LiFeAs \cite{Nag2016} great care is required in selecting the position for STS as to ensure that no artifacts from such hidden ``second-layer'' defect bound states affect the taken data.

\begin{figure}[ht]
\includegraphics*[width=\linewidth]{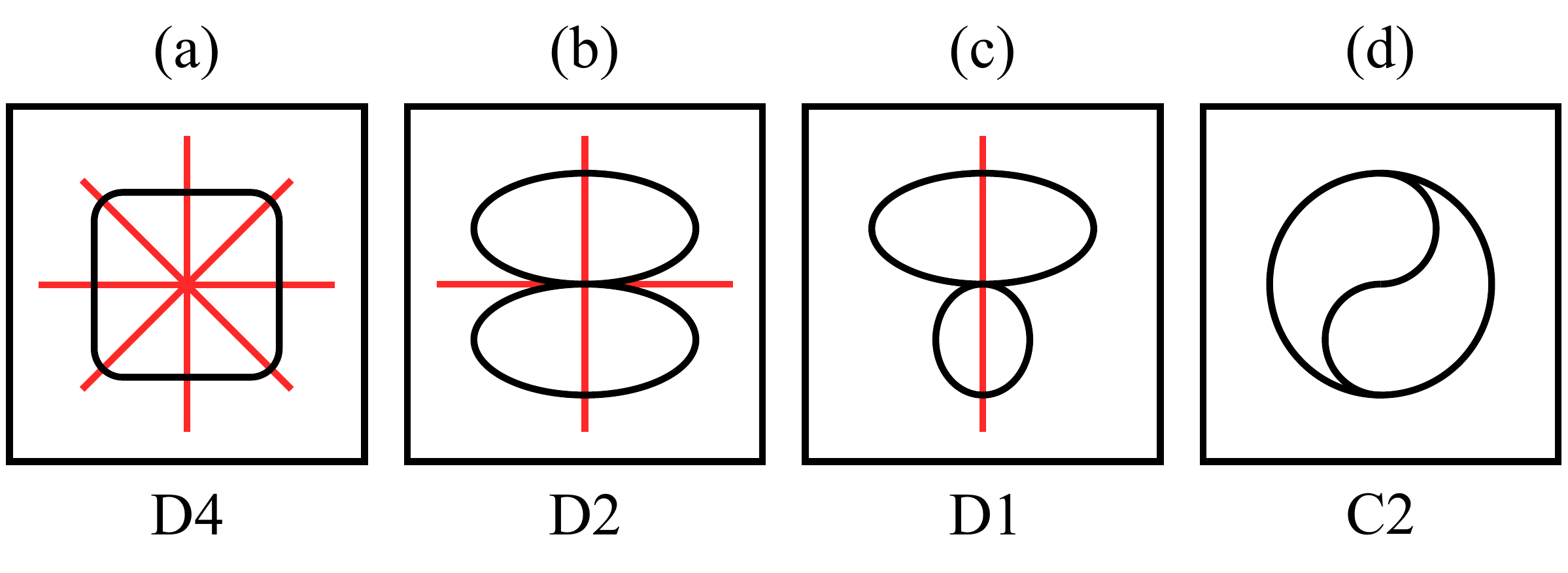}
\caption{Schematic illustration of the Sch\"onflie\ss\ notation for characterizing the observed defects. Symmetry axes are indicated as straight lines. (a-c) $n$-fold rotation-reflections $D_n$ with $n=4$, 2, 1. (d) Two-fold rotation $C_2$.}
\label{schoenfliess}
\end{figure}

\subsection{Defect types}
Upon inspection of the observed defects we were able to identify seven qualitatively different defect types. Fig.~\ref{all_types_of_defects} depicts representative topographic data including a height profile through the center of the defects.
In order to classify the different defect types that are present in Fig.~\ref{topo_overview}a we follow the approach of Grothe et al. \cite{Grothe2012} and use the Sch\"onflie\ss\ symmetry notation as indicated in Fig.~\ref{schoenfliess}. Within this approach and considering their topographic appearance and lattice position the defects can be grouped into four different sets. 

The first set is composed of the two well-confined defect types shown in Fig.~\ref{all_types_of_defects}(a) and (b). These defects appear as topographic holes and protrusions, respectively. These defects have been observed to be highly mobile, i.e., during the measurement they are moved over the surface by the tunneling tip, or are even removed completely. Due to this high mobility and the spatially very confined electronic influence on the environment we assign these defects to one missing and one extra Li atom respectively. Accordingly, we label these defects as ``Li-$D_4$-1'' and ``Li-$D_4$-2''.

Defects with  $D_2$ symmetry as shown in Fig.~\ref{all_types_of_defects}(c) and (d) form the second set of defects and have the highest abundance of all defects. Their mirror symmetry axes are parallel to the lattice constant $a=3.7914$~\AA. The LDOS modifications for the two defect types yield similar but distinct hourglass-like shapes in the topography, where the maximum height difference of about 1.15~\AA\ of the defect in Fig.~\ref{all_types_of_defects}(c) is about two times higher than that in Fig.~\ref{all_types_of_defects}(d). The comparison with the $c$-axis projection of the LiFeAs lattice shown in Fig.\ref{all_types_of_defects}(h) reveals that only Fe atoms possess a $D_2$ symmetry in the lattice. Thus, we assign the two defects in Fig.~\ref{all_types_of_defects}(c) and (d) to defects at the Fe site, consistent with the assignment by Grothe et al.~\cite{Grothe2012} and label them ``Fe-$D_2$-1''  and ``Fe-$D_2$-2'', respectively.

The third set consists of another defect type with $D_4$ symmetry  with a topographic extension about 4~nm that is clearly larger than that of ``Li-$D_4$-1'' and ``Li-$D_4$-2'' (cf. Fig.~\ref{all_types_of_defects}(e)). Both Li and As positions are compatible with the $D_4$ symmetry. However, the apparent large extended influence on the LDOS suggests this defect to be located on an As- rather than a Li-site, because the As $4p$ states partially hybridize with the Fe $3d$ states, in stark contrast to Li \cite{Lankau2010}. We therefore assign the defect to the As-site and label it ``As-$D_4$''.

Defects with a symmetry that is incompatible with that of any cite in LiFeAs form the fourth set of observed defects. The defect in Fig.~\ref{all_types_of_defects}(f) has $D_1$ symmetry with a symmetry axis that is rotated by 45$^\circ$ with respect to the lattice constant, and an enhancement of the LDOS in the upper quarter of the defect which is clearly visible in height profile, too. One might speculate that the origin of such defects lies in dimer or trimer configuration of defects. However, a further clarification of their origin seems unfeasible with the present data. Similar holds for chiral defects with $C_2$ symmetry shown in Fig.~\ref{all_types_of_defects}(g). These defects are very rare with both chiralities appearing in the data.

Of course, the abundance of above discussed defects is sample dependent as they are connected with the chemical composition and sample purity. For example, in a representative studied crystal, the defect concentration is found to be 0.4 $\pm$ 0.08$\%$ per unit cell. This statistics results from a total of 2858 measured defects on 6 different topography images with total area of 11000 nm$^2$. Fe-$D_2$-1 defects were observed most frequently (about 38\%), whereas both Fe- $D_2$-2 and $D_1$ defects occurred with an abundance of about 21\% each. The other defects mentioned above altogether give rise to about 20\% of all defects.

\begin{figure*}[ht]
	\centering
		\includegraphics[width=\textwidth]{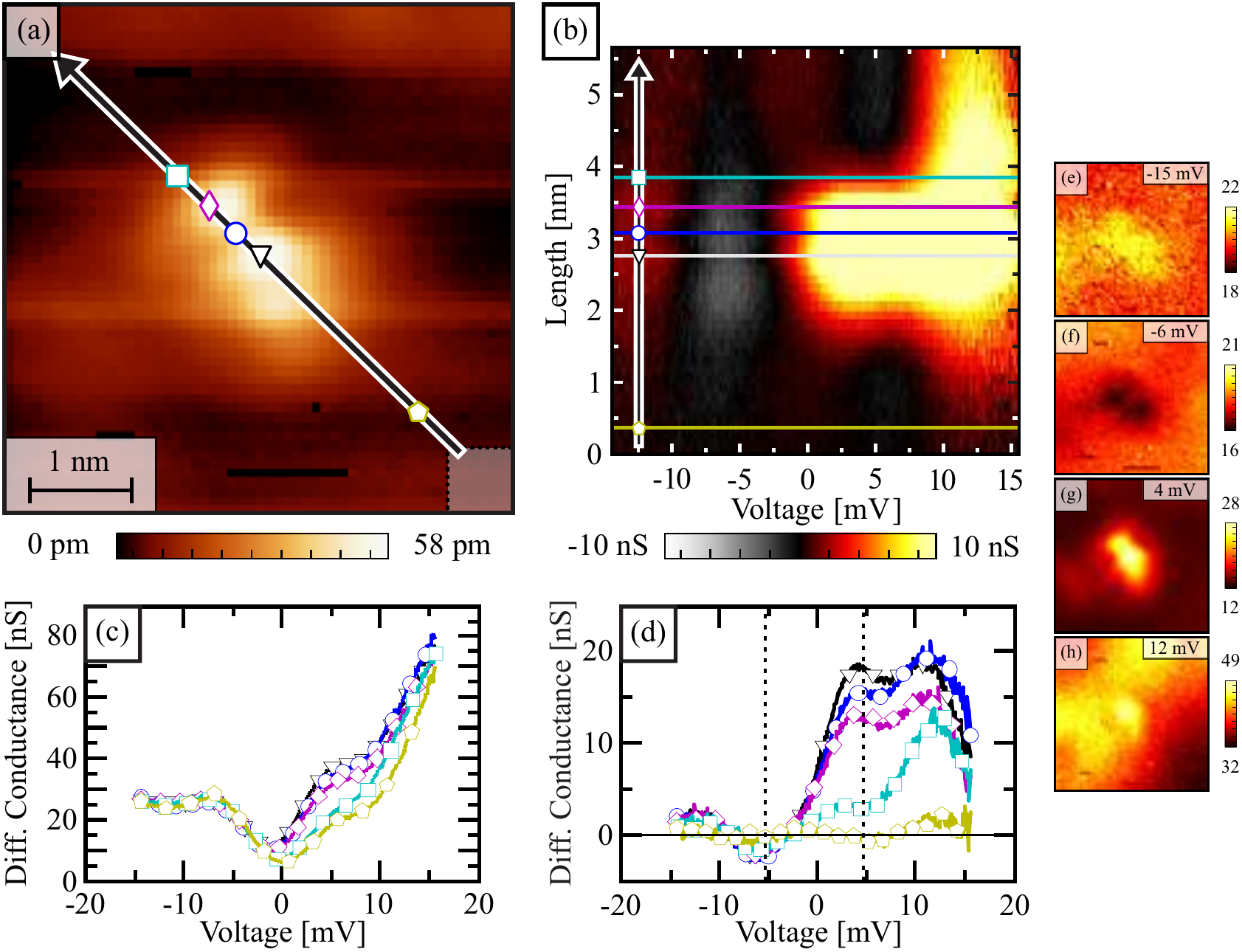}
\caption{Spectroscopy map of a Fe-$D_2$-1 defect with a FOV of 5 nm $\times$ 5 nm at 5~K. a) Topography image ($\textit{U}_{t}$ = -15 mV; $\textit{I}_t$ = 300 pA). A full spectroscopy map has been taken at the same time with 56 pixels $\times$ 56 pixels lateral resolution where each $dI/dU$ spectrum has been taken between $\pm$15 mV with a resolution of 0.1 mV for each pixel. The shaded square box at the lower right corner indicates the area where $dI/dU_\mathrm{ref}$ has been determined (see text). b) Spectra along the arrow in (a) after subtracting $dI/dU_\mathrm{ref}$ as a function of distance. c) Single point spectra according to symbols along the arrow in (a). d) Point spectra of (c) after subtracting $dI/dU_\mathrm{ref}$. e)-h): $dI/dU$ maps at -15 mV, -6 mV, 4 mV, 12 mV.}
\label{Fe_D2_1_defect}
\end{figure*}

\subsection{Defect spectroscopy}
In order to gain further information on how the defects affect the superconducting state of LiFeAs, spectroscopic maps have been measured on isolated and stable Fe-$D_2$-1, Fe-$D_2$-2, $D_1$, and As-$D_4$ defects. 

\begin{figure*}[ht]
	\centering
		\includegraphics[width=\textwidth]{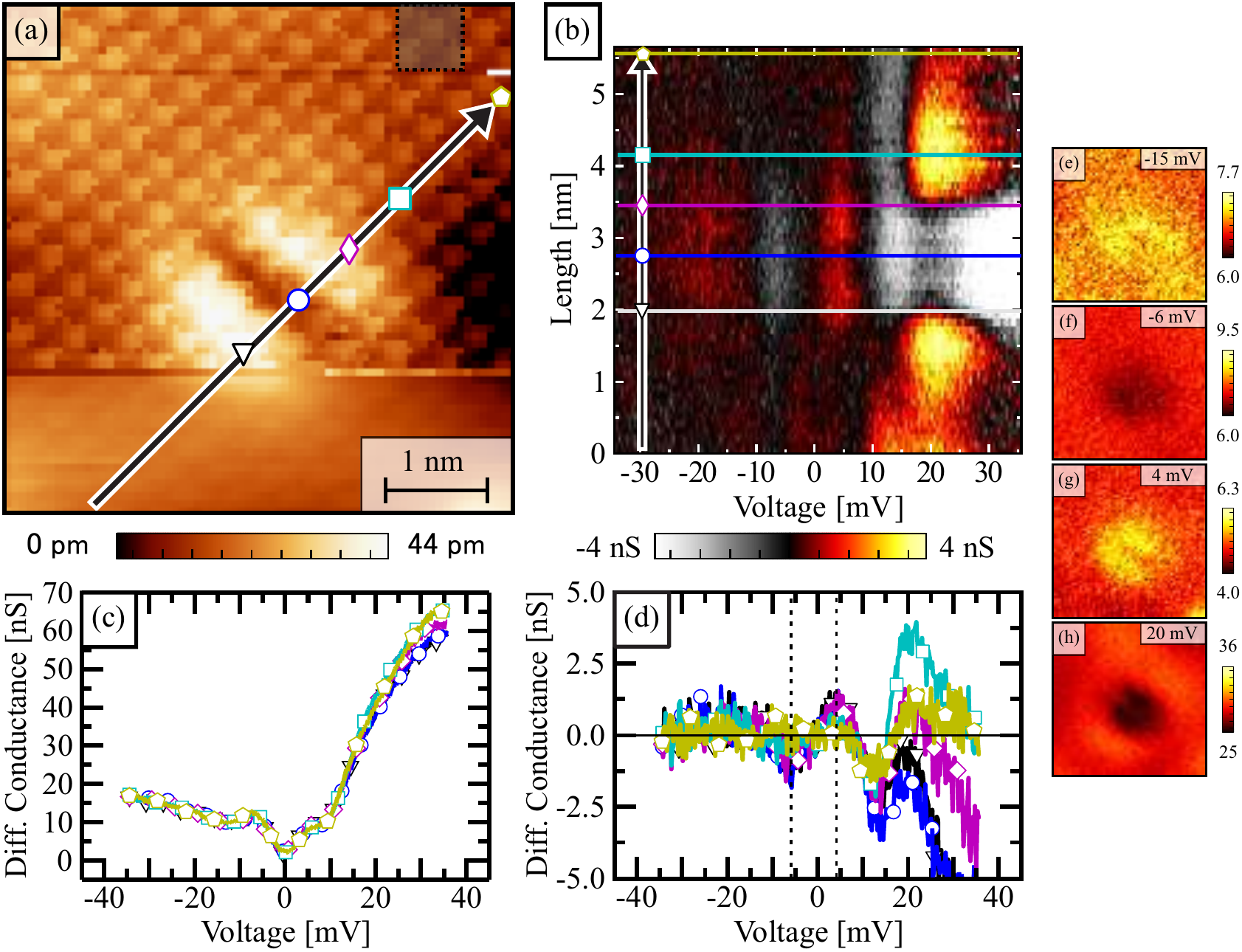}
\caption{Spectroscopy map of a Fe-$D_2$-2 defect with a FOV of 5 nm $\times$ 5 nm at 5~K. a) Topography image ($\textit{U}_{t}$ = -35 mV; $\textit{I}_t$ = 400 pA). A full spectroscopy map has been taken at the same time with 70 pixels $\times$ 70 pixels lateral resolution where each $dI/dU$ spectrum has been taken between $\pm$25 mV with a resolution of 0.25 mV for each pixel. The shaded square box at the right upper corner indicates the area where $dI/dU_\mathrm{ref}$ has been determined (see text). b) Spectra along the arrow in (a) after subtracting $dI/dU_\mathrm{ref}$ as a function of distance. c) Single point spectra according to symbols along the arrow in (a). d) Point spectra of (c) after subtracting the $dI/dU_\mathrm{ref}$. e)-h): $dI/dU$ maps at -15 mV, -6 mV, 4 mV, 20 mV.}
\label{Fe_D2_2_defect}
\end{figure*}

\subsection*{Fe-$D_2$-1 defects}
Figure~\ref{Fe_D2_1_defect}(a) shows a topographic image of a Fe-$D_2$-1 defect in a 5~nm~$\times$~5~nm FOV on which at a full $dI/dU$ spectrum has been taken between $\pm15$~mV. The square at the bottom-right corner indicates an area significantly far away from the defect from which the average spectrum $dI/dU_\mathrm{ref}$ is used as a reference for further analysis. The long arrow depicts a trace along which all corresponding $dI/dU$ spectra are shown in Fig.~\ref{Fe_D2_1_defect}(b) in a false-color plot, after subtraction of $dI/dU_\mathrm{ref}$. From these data it is already apparent that the impact of the defect on the superconducting state is very asymmetric in energy. More specifically, at positive bias voltage the data reveal a significant enhancement of the LDOS, whereas at negative bias voltage the influence on the LDOS is much weaker; at about $-6$~mV and at $U_t\lesssim -10$~mV a slight reduction and a slight enhancement are present, respectively. Upon closer inspection at positive bias voltage one can recognize enhanced LDOS between zero and about $8$~mV, which has a lateral width of around 2~nm along the trace, about the defect's center. Interestingly, at higher bias voltage, centered around $\sim12$~mV, a much wider lateral spread of enhanced LDOS of the order of 5~nm becomes apparent. 

In order to investigate this further, we inspect individual point spectra along the trace as shown in Fig.~\ref{Fe_D2_1_defect}(c). As in Fig.~\ref{Fe_D2_1_defect}(b), these data show a significantly enhanced LDOS at positive bias voltage, whereas along the trace the LDOS remains practically unchanged at negative bias voltage. Fig.~\ref{Fe_D2_1_defect}(d) shows the same spectra after subtracting the reference spectrum $dI/dU_\mathrm{ref}$. These curves reveal spectroscopically that two distinct peak-like features located at $\sim4$~mV and $\sim 12$~mV  are present and are most pronounced at the center of the defect. Furthermore, these curves are consistent with the earlier statement of slightly reduced and enhanced LDOS at about $-6$~mV and at $U_t\lesssim -10$~mV, respectively. A further investigation on Fig.~\ref{Fe_D2_1_defect}(b) reveals that these faint features also possess a lateral width of around 2~nm along the trace, like the feature at $\sim4$~mV. 

In Fig.~\ref{Fe_D2_1_defect}(e) to (h) we investigate two-dimensional $dI/dU$-maps for exploring the spatial distribution of the most prominent spectral features at $U_t\lesssim -10$~mV, at about $-6$~mV, $4$~mV, and $12$~mV. It becomes apparent from these maps that the former three features share a similar spatial distribution, viz. an elongated structure (2:1 ratio of axes) oriented along the axis which connects the two lobes of the visual shape of the defects in Fig.~\ref{all_types_of_defects}(c) and Fig.~\ref{Fe_D2_1_defect}(a). Remarkably, as is evident from Fig.~\ref{Fe_D2_1_defect}(h), the enhanced LDOS at $\sim12$~mV is most pronounced at the center of the defect as well at two extended lobes which are 90$^\circ$ rotated with respect to the topographic long axis of the defect, and which are separated about 4.5~nm apart.

We mention that the observed spectral feature at about 4~meV is roughly consistent concerning energy position and spatial distribution of the ``Fe- $D_2$-1''-defect resonance at about $3$~mV described by Grothe et al. \cite{Grothe2012}. Furthermore, these authors reported also depleted LDOS at negative bias voltage, however much more pronounced as in our data. More interesting is that our features at $U_t\lesssim -10$~mV and in particular  at $\sim12$~mV are not observed in Ref.~\cite{Grothe2012}. A possible reason for this apparent difference might be the fact that our stabilization voltage in the spectroscopy maps was negative, in contrast to those used by the aforementioned authors.

\begin{figure*}[ht]
\centering
\includegraphics[width=\textwidth]{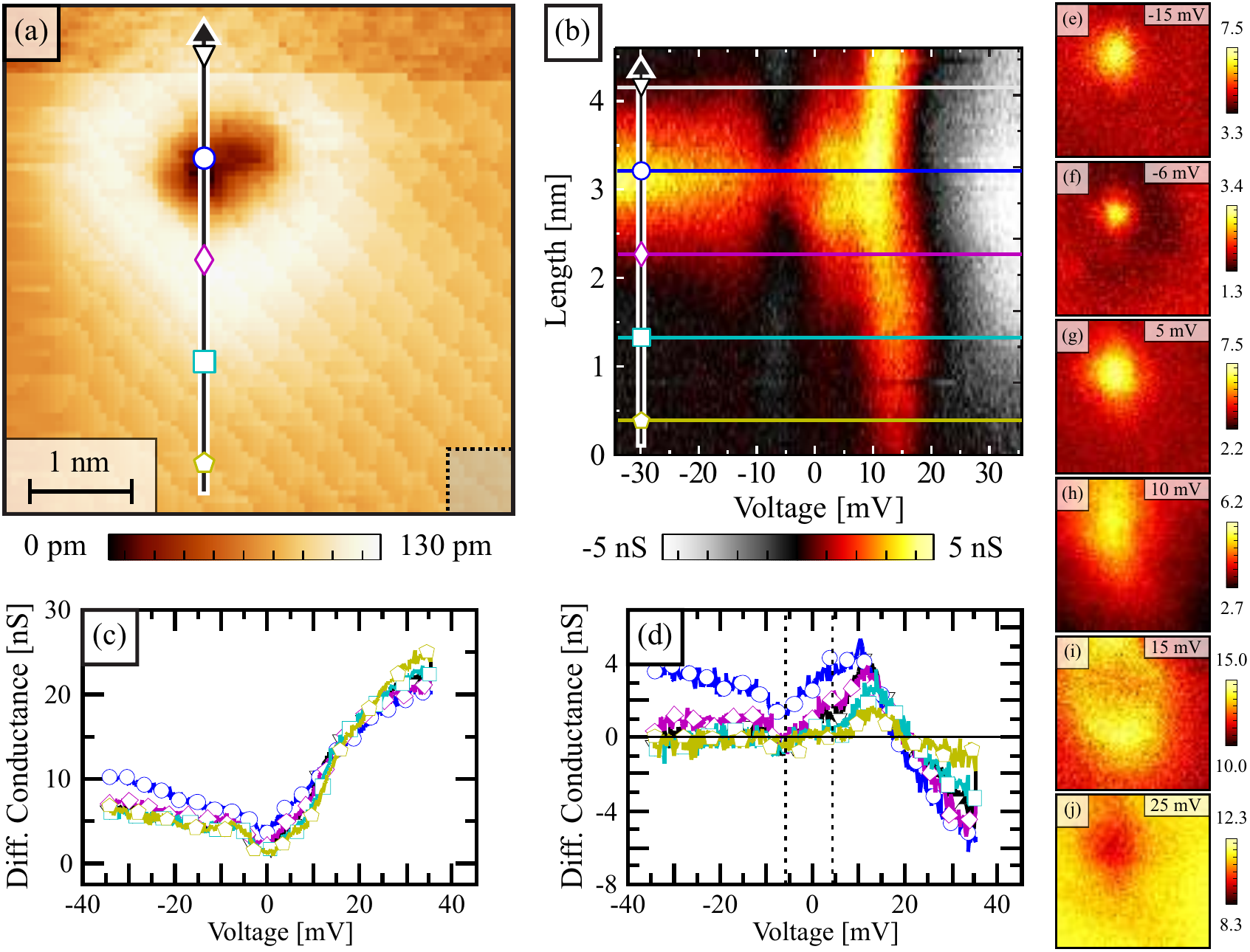}
\caption{Spectroscopy map of a As-$D_4$ defect with a FOV of 5 nm $\times$ 5 nm at 5~K. a) Topography image ($\textit{U}_{t}$ = +35 mV; $\textit{I}_t$ = 500 pA). A full spectroscopy map has been taken at the same time with 80 pixels $\times$ 80 pixels lateral resolution where each $dI/dU$ spectrum has been taken between $\pm$35 mV with a resolution of 0.25 mV for each pixel. The shaded square box at the lower right corner indicates the area where $dI/dU_\mathrm{ref}$ has been determined (see text). b) Spectra along the arrow in (a) after subtracting $dI/dU_\mathrm{ref}$ as a function of distance. c) Single point spectra according to symbols along the arrow in (a). d) Point spectra of (c) after subtracting $dI/dU_\mathrm{ref}$. e)-j): $dI/dU$ maps at -15 mV, -6 mV, 5 mV, 10 mV, 15 mV, 25 mV.}
\label{As_D4_defect}
\end{figure*}

\subsection*{Fe-$D_2$-2 defects}
Fig.~\ref{Fe_D2_2_defect}(a) shows a topographic map of a Fe-$D_2$-2-defect which has been taken during spectroscopic investigation. In the lower part of the image a changed tip state becomes apparent from the obvious much weaker spatial resolution. We stress that this change of the tip state has no noticeable effect on the spectroscopic data as becomes apparent from $dI/dU$-maps discussed further below. In a similar fashion as for the Fe-$D_2$-1 defects described above, we analyze at Fig.~\ref{Fe_D2_2_defect}(b) the spectral evolution along a trace through the two topographic lobes of the defect as indicated in Fig.~\ref{Fe_D2_2_defect}(a). At first glance, some similarity to spectral features of the Fe-$D_2$-1 defects are apparent: at  about $-6$~mV ($4$~mV) a slight depletion (enhancement) of the LDOS with respect to $dI/dU_\mathrm{ref}$  is visible with roughly the same extension along the trace (about 2~nm) as the corresponding features in Fig.~\ref{Fe_D2_1_defect}(b). Furthermore, a slightly enhanced LDOS at  $U_t\lesssim -15$~mV with similar spatial extension resembles the observed feature at $U_t\lesssim -10$~mV in the Fe-$D_2$-1 defect. However, instead of a further enhanced LDOS at about 12~mV, these defects exhibit \textit{depleted} LDOS around this energy. Only at $U_t\gtrsim 15$~mV a significant enhancement of the LDOS is visible. Interestingly, however, this enhancement is apparent only about 1~nm away from the defect center. At lower distances it remains depleted. 

Fig.~\ref{Fe_D2_2_defect}(c) shows the selected point spectra along the trace. Very clearly, all defect-induced effects on the spectra are very weak. In order to be able to discern the described features of Fig.~\ref{Fe_D2_2_defect}(b), the difference spectra with respect to $dI/dU_\mathrm{ref}$ as shown in Fig.~\ref{Fe_D2_2_defect}(d) are necessary. The data allow to identify maximum spectral changes at about $-6$~mV, $4$~mV, and $20$~mV. We therefore plot in Fig.~\ref{Fe_D2_2_defect}(e)-(h) the two-dimensional $dI/dU$-maps around the defect.
First of all, all these maps, which have been recorded simultaneously with the topographic map in Fig.~\ref{Fe_D2_2_defect}(a), show no impact of the changed tip state, as mentioned already above. It is obvious from the data at $-15$~mV that practically no spatially dependent spectroscopic influence occurs at this energy. This is different at the other three energy values. At $-6$~mV and $4$~mV a weak but clear depletion and enhancement of the LDOS is visible which is quite confined to a roundish structure of about 2~nm diameter. At $12$~mV the spectroscopy map attains a similar   $D_2$ symmetric shape as in the topographic data, however with inverted contrast and a clear almost circular ridge of enhanced LDOS at about 1.5~nm away from the defect's center is apparent.

All together the observed spectral features resemble significantly those of the Fe-$D_2$-1-defect with the strong difference that here all observed spectral features are very weak. This suggests either that the impurity atoms that give rise to the two different appearances of Fe-$D_2$-defects are different, or that both appearances belong to the same impurity type, however in different FeAs-planes. It is not possible on basis of the data to rule out one of the possibilities. Yet, the latter possibility seems more likely, given that the strongest difference between both defect types is mainly the intensity.

\subsection*{As-$D_4$ defects}
Figure~\ref{As_D4_defect}(a) shows the topographic measurement obtained during spectroscopically mapping out one of the As-$D_4$ defects. It is characterized by a ring-like enhancement of the topography with a diameter of about 2~nm, with its center appearing deeper than the surrounding atoms. The investigation of the trace through the defect shown in Fig.~\ref{As_D4_defect}(b) already reveals a significantly different impact of this defect type on the LDOS as compared to the Fe defects discussed above. The main difference is a significantly enhanced LDOS at negative bias voltages with a width of about 1.5~nm which is practically unchanged between about $-10$~mV and $-35$~mV. In Fe defects, the enhancement at negative bias voltages is in contrast almost negligible. Between $-10$~mV and about $-6$~mV, the width of this enhancement reduces to about 0.5~nm. At further increased  bias voltage, the spatial width of the structure with enhanced LDOS increases rapidly up to about 3~nm at 5~mV and afterwards even more, exceeding the FOV. Based on the $D_4$ symmetry of the defect, we estimate a total diameter of about 7~nm for the enhanced LDOS structure at 15~mV. A further increase of the bias voltage leads to a depletion of the LDOS at the defect's center, which acquires a diameter of about 2.5~nm at 25~mV and more at higher $U_t$.

The point spectra in Figures~\ref{As_D4_defect}(c) and (d) reflect this behavior very clearly. At the center of the defect the LDOS is enhanced in the entire regime $U_t\lesssim17$~mV with a minimum at about $-6$~mV, a shoulder and a peak at about 5~mV and 10~mV. A strong decrease of the LDOS at higher $U_t$ is apparent which leads to depleted LDOS at $U_t\gtrsim17$~mV. In Figures~\ref{As_D4_defect}(e) to (j) we plot the two-dimensional $dI/dU$-maps around the defect at the selected energies $-15$~mV, $-6$~mV, 5~mV, 10~mV and 25~mV. These maps reveal a almost circularly enhanced LDOS at $U_t\lesssim5$~mV with a minimum lateral diameter at $-6$~mV. At 10~mV and 15~mV the shape of the structure with enhanced LDOS appears elongated and seemingly develops two lobes in the latter case. At 25~mV the map reveals again an almost circular structure, now with a depletion at its center. While the overall spatial development of the spectroscopic defect appearance is in accordance with the data in Figures~\ref{As_D4_defect}(b) and (h), underpinning the overall data consistency, the seemingly broken $D_4$ symmetry is puzzling. We attribute it to imperfect tunneling conditions rather than a true broken symmetry. Nevertheless, the issue appears worthwhile to be readdressed in future experiments.

\subsection*{$D_1$ defects}
\begin{figure*}[ht]
\centering
\includegraphics[width=\textwidth]{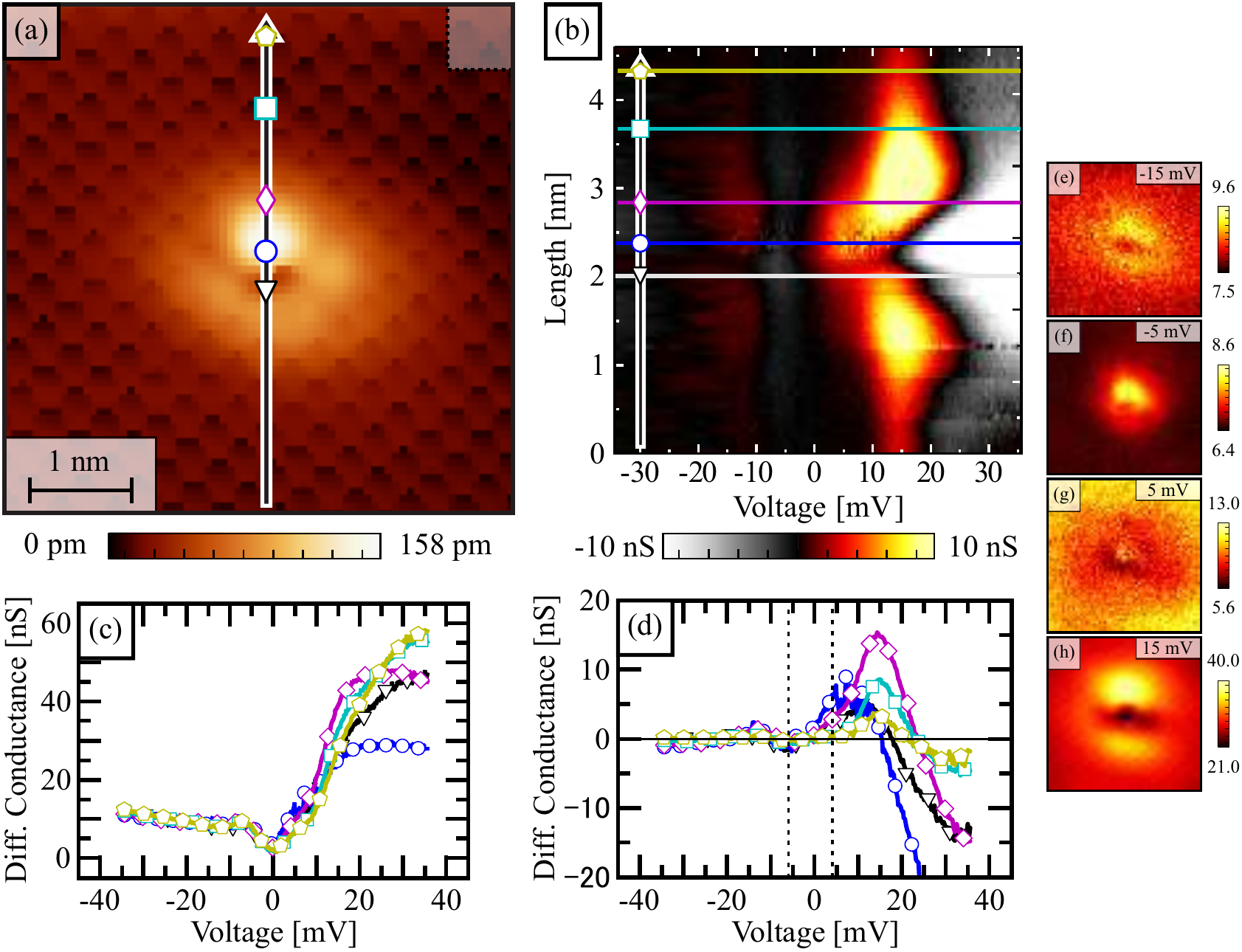}
\caption{Spectroscopy map of a Fe-$D_1$ defect with a FOV of 5 nm $\times$ 5 nm at 5~K. a) Topography image ($\textit{U}_{t}$ = -35 mV; $\textit{I}_t$ = 300 pA). A full spectroscopy map has been taken at the same time with 71 pixels $\times$ 71 pixels lateral resolution where each $dI/dU$ spectrum has been taken between -15 mV to +35 mV with a resolution of 0.25 mV for each pixel. The shaded square box at the upper right corner indicates the area where $dI/dU_\mathrm{ref}$ has been determined (see text). b) Spectra along the arrow in (a) after subtracting $dI/dU_\mathrm{ref}$ as a function of distance. c) Single point spectra according to symbols along the arrow in (a). d) Point spectra of (c) after subtracting $dI/dU_\mathrm{ref}$. e)-h): $dI/dU$ maps at -15 mV, -5 mV, 5 mV, 15 mV.}
\label{Fe_D1_defect}
\end{figure*}

$D_1$ and $C_2$ type defects belong to those defect types with a measured spatial symmetry which is incompatible with  the lattice symmetry. It is thus impossible to determine their lattice position on basis of the present data. As already mentioned above, thinkable origins of these defects are dimer or trimer configurations of defects. Due to these difficulties in further assigning the defect nature, we exemplarily focus here only on spectroscopic data of the ``$D_1$'' defects. The topography image in Fig.~\ref{Fe_D1_defect}(a) suggests a mirror symmetry axis that is rotated about $45^{\circ}$ with respect to the lattice constants. Remarkably, the  maximum height of the topography is located outside of the center of the defect. The spectroscopic analysis along the trace depicted in Fig.~\ref{Fe_D1_defect}(b) reveals similarities as compared to the Fe-defects discussed above, as there are no pronounced features at negative $U_t$, except a slightly enhanced LDOS at the ``center'' of the defect. Upon increasing $U_t$, a significant enhancement rapidly develops in the form of two lobes with asymmetric size and but similar shape dispersing outward the defect. At  $U_t\gtrsim15$~mV a depletion of the LDOS develops.

The point spectra in Fig.~\ref{Fe_D1_defect}(c) and (d) corroborate this observation. In particular, in Fig.~\ref{Fe_D1_defect}(d) one can observe a slightly enhanced LDOS at around $-15$~mV in the center of the defect. At positive bias voltage the point spectra become very position dependent. Spectra in the ``lower'' part of the trace shown in Fig.~\ref{Fe_D1_defect}(a) possess a broad peak between about 5~mV and 10~mV followed by a strong decrease, whereas for those in the upper part the broad peak is around 15~mV. These overall $D_1$ symmetric properties are well reflected also in the spectroscopic maps in Fig.~\ref{Fe_D1_defect}(e) to (h).

\section{Discussion}

LiFeAs possesses a complicated multiorbital band structure \cite{Ahn2014a}. Thus one can expect, in principle, a rather complicated multi-peak structure of impurity bound sates in the superconducting state \cite{Kariyado2010}. It is important to note, that in the tunneling conditions applied here, i.e., at about 5~K measurement temperature, the energy resolution is $\Delta\varepsilon\gtrsim3.5k_BT\approx 1.5$~meV. Thus, given the sizes of the known superconducting gaps in this compound of $\Delta_1 \approx 6$~meV and $\Delta_2\approx 3.5-4$~meV \cite{Chi2012,Hanaguri2012,Borisenko2012}, one can expect to be able to resolve coarse structures of impurity states only, without any fine-structure. 
Thus, only the most salient features in the above defect spectroscopy can be considered for further interpretation of the data. A priori, it is unclear to what extent the observed natural impurities can be classified in terms of non-magnetic and magnetic defects. However, among the observed defects, concerning their overall impact on the local LDOS they seem to represent two very distinct classes: Generally, the Fe-defects and the $D_1$ defects have in common that there is only a very weak impact of the defect at negative bias voltages. Furthermore, they share multiple pronounced features of enhanced LDOS that occur at positive bias: a low-energy feature at about 4~mV and salient features of enhanced LDOS at $\gtrsim10$~mV. The former corresponds only roughly with the size of the small gap $\Delta_2$. The bound state's position is however located at somewhat larger energy, which suggests it to be connected rather with the larger gap $\Delta_1$. The latter defect states  at $U_t\gtrsim10$~mV are clearly at larger energies than any known superconducting gaps in LiFeAs. One might speculate that even larger superconducting gaps than the observed ones of the order of 10~meV exist in this compound or attribute the high-energy features to defect-induced states unrelated to superconductivity. A closer investigation of the gap structure e.g. by ultra-low temperature STS studies or ARPES experiments which exceed the current resolution as well as normal state defect spectroscopy could clarify this point.

On the other hand, the As-$D_4$ defects possess a spectral fingerprint that is drastically different, since the LDOS ins strongly enhanced also at negative bias voltages, in addition to features at about 4~mV and 10~mV. This suggests that the nature of this defect type is completely different from that of the Fe-defect and the $D_1$ defect.

At present, it appears impossible to assign these defect type groups a distinct character in terms of potential or magnetic scattering,  which could serve as basis to conclude the superconducting order parameter from the spectroscopic data. However, a possible route towards achieving this is to theoretically model the real space impact of particular defect  types with respect to defect-induced local magnetism \cite{Grinenko2011,Gastiasoro2014a,Gastiasoro2014}, or to study their magnetic properties by means of spin-polarized STM \cite{Wiesendanger2009}. In any case, it seems desirable to perform high-resolution (i.e. ultra-low temperature) STS studies on these defects, as well as to develop a specific model for impurity-induced bound states specifically for LiFeAs, for extracting more solid information about the order parameter in this intriguing compound.

\section{Summary}

We have investigated the topographic appearance and the impact of natural impurities in stoichiometric LiFeAs on the local density of states (LDOS). Our findings reveal not only a clear sign dependence of the tunneling voltage $U_t$ on the topographic appearance, i.e., defects appear spatially more extended at positive $U_t$, we also find that for most defect types the strongest enhancement of the LDOS is found at positive $U_t$. More specifically, for Fe-defects with $D_2$ symmetry the LDOS is negligibly influenced at $U_t<0$ whereas a peak-like enhancement of the LDOS at about 4~mV is present and indicative of an impurity-induced bound state. Further peak-like enhancements of the LDOS appears at $U_t\gtrsim12$~mV which is at energies larger than known superconducting gaps of LiFeAs. In contrast, As-defects with $D_4$-symmetry lead to strong enhancement at both positive and negative $U_t$. Here, the enhancement of the LDOS occurs over an extended energy range which clearly exceeds the known superconducting gaps at both polarities. Furthermore, a distinct enhancement of the LDOS is present at about 4~mV and 12~mV, where the latter shows the largest extension.

\section{Acknowledgement}
This work has been supported by the Deutsche Forschungsgemeinschaft (DFG) through the Priority Programme SPP1458 (Grant HE3439/11 and BU887/15-1), and the Graduate School GRK1621.  Furthermore, this project has received funding from the European Research Council (ERC) under the European Union’s Horizon 2020 research and innovation programme (grant agreement No 647276 -- MARS -- ERC-2014-CoG). S.W. acknowledges funding by DFG under the Emmy-Noether program (Grant No. WU595/3-3).

\providecommand{\WileyBibTextsc}{}
\let\textsc\WileyBibTextsc
\providecommand{\othercit}{}
\providecommand{\jr}[1]{#1}
\providecommand{\etal}{~et~al.}

\end{document}